\documentclass[aps, twocolumn, showpacs,superscriptaddress]{revtex4-1}

\usepackage{amssymb}
\usepackage{amsmath}
\usepackage[T2A]{fontenc}
\usepackage{graphics}
\usepackage{graphicx}
\usepackage[dvips]{color}

\begin{document}

\title{Spin Seebeck effect and phonon energy transfer in heterostructures  containing layers
of a normal metal and a ferromagnetic insulator}

\author{A. I. Bezuglyj}\email{bezuglyj@kipt.kharkov.ua}
\affiliation{\it National Science Center "Kharkov Institute of Physics and Technology", 1,
Akademicheskaya St., Kharkov, 61108, Ukraine\\} \affiliation{\it Kharkov National University, 4,
Svobody Sq., Kharkov, 61022, Ukraine }

\author{V. A. Shklovskij}\email{shklovskij@univer.kharkov.ua}
\affiliation{\it Kharkov National University, 4, Svobody Sq., Kharkov, 61022, Ukraine }

\author{V. V. Kruglyak}
\affiliation{\it School of Physics and Astronomy, University of Exeter, United Kingdom }

\author{R. V. Vovk}
\affiliation{\it Kharkov National University, 4, Svobody Sq., Kharkov, 61022, Ukraine }

\date{\today}

\begin{abstract}
In the framework of the kinetic approach based on the Boltzmann equation for the phonon
distribution function,  we analyze phonon heat transfer in a heterostructure containing a layer of
a normal metal ($ N $) and a layer of a ferromagnetic insulator ($ F $). Two  realistic methods
for creating a temperature gradient in such a heterostructure are considered: by heating of the
$N$-layer by an electric current and by placing the $N/F$-bilayer between massive dielectrics with
different temperatures. The electron temperature $ T_e $ in the $ N $-layer and the magnon
temperature $ T_m $ in the $ F $-layer are calculated. The difference in these temperatures
determines the voltage $ V_{ISHE} $ on the $ N $-layer in the Seebeck spin effect regime. The
dependence of $ V_{ISHE} $ on the bath temperature  and on the thickness of the $ N $ and $ F $
layers is compared with the available experimental data.
\end{abstract}

\maketitle

\section{Introduction}\label{s0}

In recent years, the relation between the processes of spin and heat transport, that is, spin
caloritronics, has been of great interest \cite{Bauer2010, Bauer2012b, Bauer2012, Boona2014}. This
interest is largely due to the recent observation of the spin Seebeck effect (SSE) in
heterostructures containing layers of a normal metal ($ N $)  and a ferromagnetic insulator ($ F
$), where heat is transferred mainly by phonons  \cite{Uchida2008, Uchida2010, Uchida2010-1,
Agrawal-APL2014, Agrawal-PRB2014, Schreier-PRB2016}. Similar to the conventional Seebeck effect,
when an electron current emerges as a result of a temperature gradient, the temperature gradient
in the spin Seebeck effect generates a spin current. Since it is not yet possible to directly
measure the spin current experimentally, a two-layer ferromagnet/normal metal heterostructure ($ F
/ N $) is used to detect the SSE. In such a structure, the spin current from the $ F $-layer is
injected into the $ N $-metal, where it induces the experimentally observed voltage $ V_ {ISHE} $
due to the inverse spin Hall effect (ISHE).\cite{Azevedo2005}

Most SSE experiments are carried out in the longitudinal (LSSE) geometry, in which the temperature
gradient $ \nabla T $ and the spin current $ {\bf J}_s $ are parallel  to each other and are
oriented perpendicular to the $ F / N $ interface.\cite{Guo-Cramer_PhysRev2016,
Schreier-Kamra-PhysRev2013} In this geometry, to exclude the anomalous voltage in the $ F $-layer
(due to the Nernst effect \cite {Huang2011}),  the $ F $-layer must be nonconducting, i.e. a
magnetic insulator. Currently, in most of the LSSE experiments, yttrium-iron garnet (YIG) is used
as a magnetic insulator, and the $ N $-layer is made of metals with strong spin-orbit interaction,
such as platinum (Pt) or gold (Au).

It is important for us that, in the LSSE experiment, the temperature gradient can be realized in
different ways. If thermal sources and sinks with different temperatures are used at the
boundaries of the $ F / N $-sample, then a good thermal contact with such thermal reservoirs
(thermostats) is needed to create a large temperature gradient. \cite{Uchida2008, Uchida2010,
Uchida2010-1} When the free surface of the heterostructure is irradiated by a laser beam, the
heating occurs locally in a rather narrow region of the sample. The temperature gradients obtained
in this way can be described quantitatively only by numerical simulation of temperature profiles.
\cite{Schreier-Kamra-PhysRev2013}  In Refs. \cite{Schreier-APL2013, Wang-APL2014}, another very
simple method is presented for creating large temperature gradients perpendicular to the
$F/N$-interface: by heating the $ N $-metal in the $F/N$-bilayer by an electric current. In this
case, the $ N $ layer used to detect LSSE is simultaneously used both as a resistive heater and  a
thermometer. In this paper, we analytically consider two ways of realizing the temperature gradient
in the LSSE experiment: a) by heating the $ N $-layer with an electric current and b) by
means of two thermostats with different temperatures $T_1 \neq T_2$.

Our investigation of the phonon energy transfer in $ F / N $ heterostructures was mainly motivated
by the results from Ref. \cite{Guo-Cramer_PhysRev2016}, where the influence of the thickness and
interface of YIG films on the low-temperature increase of the electric signal in the $ N $-layer
(Pt) under the conditions of the LSSE effect is discussed. Specifically, in
\cite{Guo-Cramer_PhysRev2016}, the dependence of the LSSE response on the temperature and the
magnitude of the magnetic field was studied in a YIG single crystal and in thin YIG films with
thicknesses from 150 nm to 50 $\mu$m. When the temperature dropped to $ \approx $ 75~K, the LSSE
signal increased significantly, and its maximum shifted toward higher temperatures with a decrease
in the thickness of YIG films and shifted to low temperatures with increasing magnetic field. To
explain these features of the LSSE response, a simple phenomenological model was proposed in  Ref.
\cite{Guo-Cramer_PhysRev2016}, based on the introduction of the temperature-dependent effective
propagation length of thermally excited magnons in YIG. In addition, Ref.
\cite{Guo-Cramer_PhysRev2016} showed that the peak of the LSSE signal is significantly shifted in
temperature with the change in the contact conditions on the $F/N$-interface.

In Ref. \cite{Schreier-Kamra-PhysRev2013},  the temperature profiles for phonons, electrons and
magnons in YIG/Pt bilayers were calculated taking into account the thermal resistances of the
interfaces (see Fig. 3 in Ref. \cite{Schreier-Kamra-PhysRev2013}, where these profiles are
depicted schematically for a heterostructure, consisting of a dielectric substrate, a
ferromagnetic insulator, and a normal metal). Recall that the concept of thermal resistance at the
interface between media (usually called the Kapitza resistance) suggests that, due to the heat
flux $ Q $ flowing through the interface between media 1 and 2, a temperature jump $\Delta T = T_1
- T_2 $ occurs on the interface. The cause of the temperature jump is the reflection of some part
of the heat carriers from the boundary. As Little has shown in the well-known article
\cite{Little1959}, at low temperatures $ T \ll \Theta_D $ (where $ \Theta_D $ is the Debye
temperature) the heat flux $ Q = A (T_{1}^{4} - T_{2}^{4})$.  The value of the constant $ A $ is
determined only by the acoustic properties of the contacting substances (i.e. their densities and
sound velocities), and the heat flux across the boundary can be calculated in the framework of the
acoustic mismatch theory \cite{Little1959}. If $ \theta $ is the angle of incidence of a phonon on
the interface, then $ A $ is proportional to the transparency coefficient $ \alpha (\theta) $,
averaged over the angles $ \theta $.  In the linear in $ \Delta T $  approximation, Little's
result leads to the Newton relation $ Q = R_{th}^{-1} \Delta T $, where the Kapitza resistance $
R_{th} = (4AT^3)^{-1}$. One can see that the value of $ R_{th} $ increases strongly ($ \propto 1 /
T^3 $)  with decreasing temperature. This result corresponds qualitatively to the low-temperature
growth of the LSSE response, obtained in Ref. \cite{Guo-Cramer_PhysRev2016}.

Concerning the theoretical ideas explaining the temperature-dependent effective effective
propagation length of thermally excited magnons in YIG introduced in  Ref.
\cite{Guo-Cramer_PhysRev2016} to explain the dimensionally dependent LSSE responses, one can refer
to  recent work \cite{Shklovskij2018}. In Ref.  \cite{Shklovskij2018} it was shown that, for the
ferrodielectric-insulator interface ($F/I$) at low temperatures ($ T \ll \Theta_D $), there exists
a {\underline {size effect}}. The latter manifests itself in the dependence of the Kapitza
resistance$ R_{th} $  for thin $ F $ plates (films) on the frequency of magnon-phonon collisions,
whereas for thick plates the value of $ R_{th} $ can be described by the Little formula, which
does not contain the magnetic characteristics of the ferrodielectric. To explain  the growth of
the magnetic  contribution with decreasing thickness of the $ F $-layer, let us consider
qualitatively the mechanism of heat transfer through the $ F /I $ interface. The transfer of heat
through the interface is provided by phonons and depends on its acoustic transparency. Then on the
$ F $ side of the interface there is a transition region in which the thermal energy transferred
by the magnons is converted into a phonon flux. The thickness of this region is of the order of
the average phonon mean free path $ l_{pm} (T_m) $ with respect to their scattering by magnons ($
T_m $ is the magnon temperature). It is intuitively obvious that, if the thickness of the $ F
$-layer $ d_f \gg l_{pm} (T_m) $,  the detailed structure of this transition region is unimportant
for calculating the Kapitza resistance $F/I$ of the boundary. This corresponds to Little's usual
approach within the framework of the acoustic mismatch theory, when the magnon contribution to $
R_{th} $ is absent. However,  if $ d_f \ll l_{pm} (T_m)$ and $\alpha\sim 1$, then most  phonons
emitted by  magnons in the film leave it without interacting with the magnons, even after several
reflections from the boundaries. As a result, in contrast to the case $ d_f \gg l_{pm} (T_m) $,
the spectral distribution of the phonons emitted by the $ F $-film depends more on the
magnon-phonon interaction than on the acoustic transparency of the $ F/I $ boundary.

The paper is organized as follows. In the Introduction (Section \ref{s0}), the formulation of the
problem of the LSSE in spin caloritronics has been discussed, and some most important and relevant
experimental studies have been reviewed. These articles stimulated the authors' interest in
carrying out the kinetic calculations on phonon energy transfer in $N/F/I$ heterostructures. In
Section \ref{s1}, the problem of the temperature profile of the $N/F/I$ heterostructure is solved
for a continuous heating of the metal layer and a temperature difference between $ F $ and $ N $
layers is found for their interface, since this temperature difference determines the strength of
the LSSE-response. The limiting cases of thin and thick $ F $ and $ N $ layers are considered. In
Section \ref{s2}, the heterostructure $I_1/N/F/I_2$ is investigated, provided that the
temperatures of the insulators $I_1$ and $I_2$ are fixed. As in Section \ref{s1},  the cases of
thin and thick $ F $ and $ N $ layers are considered. In Section \ref{s3}, the theory is compared
with experiments and the main conclusions of our analytical calculations are formulated.

\section{Joule heating of the metal layer}\label{s1}

\subsection{The  problem of heat transfer in a multilayer heterostructure}\label{ss1.1}

\begin{figure}[t]
\includegraphics[width=9.5 cm]{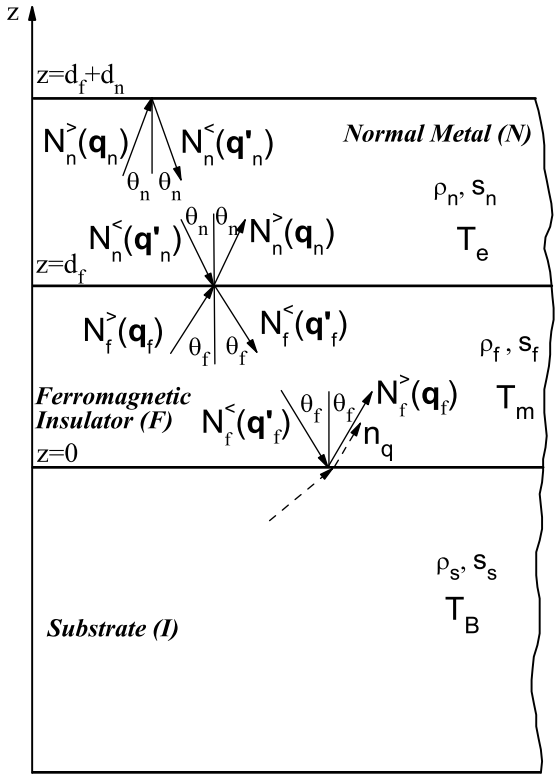}
\caption{Refraction and reflection of phonon modes at layer boundaries in the $ N / F / I $
structure. The occupation numbers of phonon states with wave vectors $ {\bf q} $ are denoted by $
N^{\lessgtr} ({\bf q}) $. The superscript $>$ marks phonons with positive $ z $-component of  $
{\bf q} $, and the superscript $<$ marks phonons with negative $ z $-component of $ {\bf q} $. The
letters $ \rho $ and $ s $ denote the densities and velocities of the longitudinal sound of the
corresponding media. $ T_e $ is the electron temperature, and $ T_m $ is the magnon temperature.
The temperature $ T_B $ is the temperature of the massive substrate, which plays the role of a
thermostat. }\label{Fig1}
\end{figure}

A layered structure $ N / F / I $ consisting of a metal layer $N$ of thickness $ d_n $ and a
ferromagnetic insulator $F$ of thickness $ d_f $ (see Fig. \ref{Fig1}) is considered. The
$ F $-layer is in direct contact with a massive dielectric substrate $ I $, which
plays the role of a thermostat with
temperature $ T_B $.

To analyze the low-temperature kinetics of electrons and magnons in this structure, we use the
following microscopic model. We assume that the energy spectrum of the electrons in the metal is
quadratic and isotropic: $ \epsilon ({\bf p}) = p^2 / (2m) $, where ${\bf p} $ is the
quasimomentum of the electron and $ m $ is its effective mass. At low temperatures, the dispersion
law of phonons can be approximated by the linear relation $ \omega_{qi} = s_{i} q_{i} $, where $
q_{i} $ is the absolute value of the phonon wave vector and $ s_{i} $ is the longitudinal sound
velocity in the corresponding medium $ (i = n, f, s) $. Here and below, we take into account only
the longitudinal acoustic branch of lattice vibrations. Taking into account the transverse
vibration modes does not change the physical picture of the energy relaxation in the layered
structure. Yet,  the inter conversion of the phonon  modes with different polarizations at the
interlayer boundaries complicates the description so much that an analytical approach becomes
formidable (see the discussion in \cite{Bezuglyi-Shklovsky-FNT2013,
Bezuglyj-Shklovskij-PhysRev2014}). The transfer of energy between electrons and phonons is a
consequence of the electron-phonon interaction, for which we use the deformation potential
approximation. The interaction of magnons with the lattice in the magnon temperature approximation
was analyzed in Ref.~\cite{Shklovskij2018}, the results of which will be used below. The energy
exchange between layers that has a phonon nature will be described in terms of the well-known
model of the acoustic mismatch. \cite{Little1959, Kaplan1979, Swartz-Pohl1989}

Suppose that the electrons in the $ N $-layer are thermalized and have a temperature $ T_e $. We
also assume that the magnons are thermalized in the $ F $-layer and their temperature is $ T_m $.
In addition, we assume that, due to the high electron and magnon thermal conductivities
\cite{Akhieser_Shishkin}, the temperatures $ T_e $ and $ T_m $ do not depend on the coordinate $ z
$ perpendicular to the layers, that is, they are constants. In contrast, the phonon distribution
functions in the $N$ and   $F$ layers depend on $ z $ and obey the kinetic equations. For the $N$
layer, we have

\begin{equation}\label{1}
 s_{nz}{{d N_n ({\bf q}_n,z)} \over{d z}} =
I_{pe}[N_n ({\bf q}_n,z)],
\end{equation}
where $ s_{nz} $ is the projection of the phonon velocity on the $ z $ axis, and $ I_{pe} $ is the
phonon-electron collision integral, which in the case of the Fermi electron distribution function
has a  simple form \cite{ShklovskiiJETP1980}:

\begin{equation}\label{2}
I_{pe}[N_n ({\bf q}_n,z )] = \nu_n (q) [n_q (T_{e}) - N_n ({\bf q}_n,z) ].
\end{equation}
Here $ n_q (T_e) = [\exp (\hbar \omega_{qn} / T_e) -1] ^{- 1} $ is the Bose distribution function
($k_B =1$). In the approximation of the deformation potential, the frequency of phonon-electron
collisions is given by

\begin{equation}\label{3}
\nu_n ={{m^2 \mu^{2}} \over{2\pi \hbar^3 \rho_n s_n}} \omega_{qn},
\end{equation}
where $ \mu $ is the deformation potential constant, which is of the order of the Fermi energy
$\epsilon_{F} $; $ \rho_n $ is the density of the metal. Note that Eq.~(\ref{3}) refers to the
case of a pure metal. In dirty metals, the dependence of $ \nu_n $ on $ \omega_{qn} $ can be
approximated by a power function with a power exponent that depends on the type of
defects.\cite{Bezuglyj-Shklovskij-JETF1997} The phonon distribution function in the magnetic layer
obeys the kinetic equation

\begin{equation}\label{1aa}
 s_{fz}{{d N_f ({\bf q}_f,z)} \over{d z}} = I_{pm}[N_f ({\bf q}_f,z)],
\end{equation}
where the collision integral of phonons with magnons has the form \cite{Shklovskij2018}

\begin{equation}\label{2aa}
I_{pm}[N_f ({\bf q}_f ,z)] = \nu_f(T_m, q) [n_q (T_{f}) - N_f ({\bf q}_f,z) ].
\end{equation}
In contrast to the phonon-electron collision frequency, which depends only on the absolute value
of the wave vector of the phonon $ q $, the frequency of phonon-magnon collisions also depends on
the temperature of the magnons $ T_m $. For the phonon-magnon collision frequency, as it was shown
in  Ref. \cite{Shklovskij2018}

\begin{equation}\label{3aa}
\nu_f(T_m, q )= D(T_m) I (T_m, q),
\end{equation}
where

\begin{equation}\label{4aa}
D(T_m)= \frac{\Theta_C}{8\pi M_f a_f s_f}\Bigl( \frac{T_m}{\Theta_C}\Bigr)^3,
\end{equation}

\begin{equation}\label{5aa}
I (T_m, q) = \int_{y_0}^\infty dy y(x+y)\Bigl( \frac{1}{e^y - 1} - \frac{1}{e^{x+y} - 1}\Bigr).
\end{equation}
We note that $ I (T_m, q) $ contains the dependence on the phonon wave vector through the
dimensionless value $ x = \hbar \omega_{qf} / T_m $. In Eq.~(\ref{4aa}), $ \Theta_C $ is the Curie
temperature, $ M_f $ is the mass of the magnetic atom, $ a_f $ is the lattice constant of the
ferromagnet. In Eq.~(\ref{5aa}), the lower limit of integration, $ y_0 = \frac {\Theta_{Df}^2}
{4T_m \Theta_C} $, accounts for that the emission of a phonon by a magnon is possible  only when
the magnon energy is greater than $ \frac {\Theta_{Df}^2} {4\Theta_C} $. Here, $ \Theta_{Df} $ is
the Debye temperature of the ferromagnet.

To formulate the boundary conditions, we turn to Fig. \ref{Fig1}, which illustrates the processes
of reflection and refraction of phonons at layer boundaries. Fig. \ref{Fig1} shows that, in the
ferromagnet $ F $ near the boundary with the substrate, the distribution function of phonons
having a positive $ z $-component of the wave vector contains two contributions. One of them is
determined by phonons coming from the substrate, and the other is due to the phonons  of the
ferromagnet, reflected from the boundary:

\begin{equation}\label{4}
N^{>}_{f} ({\bf q}_f, 0)=\alpha_{s\rightarrow f}(\theta_s) n_{q}(T_B)+\beta_{f\rightarrow
s}(\theta_f) N^{<}_{f} ({\bf q'}_f, 0).
\end{equation}
Hereinafter, the Greek letters $ \alpha $ and $ \beta $ denote the probability of passing a phonon
through the boundary between adjacent materials and the probability of reflection from the
boundary ($ \beta = 1- \alpha $). The subscripts in $ \alpha $ and $ \beta $ define the boundary.
$ T_B $ is the temperature of the substrate. Everywhere in the boundary conditions, the wave
vectors $ {\bf q} $ and $ {\bf q '} $ represent phonons that have a positive or negative
$z$-component of the wave vector: $ {\bf q} = (q_x, q_y, q_z > 0) $, $ {\bf q^{\prime}} = (q_x,
q_y, q_z<0) $. Correspondingly, $ N^{>} $ and $ N^{<} $ denote the occupation numbers of states
with wave vectors $ {\bf q} $ and $ {\bf q'} $.

The condition (\ref {4}) assumes that the phonons that have flown from the $ F $ layer to the
substrate do not return back. Such a picture is typical for single-crystal substrates with high
thermal conductivity, where phonons propagate ballistically.

In the acoustic mismatch model \cite{Little1959, Kaplan1979, Swartz-Pohl1989}, the probability of
passing the interface depends on the angle of incidence of the phonon $ \theta $ and the acoustic
impedances of the adjacent media $ Z = \rho s $ and $ Z^{\prime} = \rho^{\prime} s^{\prime} $:

\begin{equation}\label{5}
\alpha(\theta)= {\frac{4ZZ^{\prime}{\rm cos}\theta{\rm cos}\theta '}{(Z{\rm cos}\theta ' +
Z^{\prime}{\rm cos} \theta)^2}}.
\end{equation}
The angles of incidence and refraction are related by $ {\rm sin} \theta = (s /
s^{\prime}) {\rm sin} \theta^{\prime} $. Here, adjacent materials are characterized by quantities
with a prime and without it.

Conditions on the boundaries $ z = d_f $ and $ z = d_f + d_n $ are written analogously to the
relation (\ref{4}). For $ z = d_f $ we have

\begin{eqnarray}\label{6}
N^{<}_{f} ({\bf q'}_f, d_f)=\alpha_{n\rightarrow f}(\theta_n) N^{<}_{n} ({\bf q'}_n, d_f)\\
\nonumber
+\beta_{f\rightarrow n}(\theta_f) N^{>}_{f} ({\bf q}_f, d_f),
\end{eqnarray}

\begin{eqnarray}\label{7}
 N^{>}_{n} ({\bf q}_n, d_f)=\alpha_{f\rightarrow n}(\theta_f) N^{>}_{f} ({\bf q'}_f,
d_f) \\ \nonumber
+\beta_{n\rightarrow f}(\theta_n) N^{<}_{n} ({\bf q}_n, d_f).
\end{eqnarray}
It follows from Eq.~(\ref{5}) that $ \alpha_{n \rightarrow f} (\theta_n) =
  \alpha_{f \rightarrow n} (\theta_f) $.

For $ z = d_f + d_n $, the boundary condition describes the specular reflection of phonons at the
outer boundary of the metallic layer:

\begin{equation}\label{8}
N^{>}_{n} ({\bf q}_n, d_f+d_n)=N^{<}_{n} ({\bf q'}_n, d_f+d_n).
\end{equation}

\subsection{Analytical solution of the heat transfer problem in a layered
$N/F/I$-system}\label{ss1.2}

The purpose of our calculations is to find the temperature difference $ T_e - T_m $, which appears
in the expression for the voltage on a normal metal under the conditions of  the SSE
\cite{Schreier-Kamra-PhysRev2013, Xiao2010}. This difference is determined by the heat flux $ Q $
from a normal metal heated by an electric current to a cooler dielectric substrate. Since the
transfer of heat through the $ N / F / I $-system is of a phonon nature, to calculate $ T_e-T_m $,
it is necessary to solve the kinetic equations for the phonon distribution function in $ N $ and $
F $ layers and to stitch the solutions at the interfaces. The solutions of these kinetic equations
have form

\begin{equation}\label{11}
N_{i}^{>}({\bf q}_i,z)= C_{i}^{>}({\bf q}_{i}) {\rm exp}(- z/l_i)+n_{q_{i}}(T_i),
\end{equation}
for $q_z > 0$ and

\begin{equation}\label{12}
N_{i}^{<}({\bf  q}^{\prime}_i,z)= C_{i}^{<}({\bf q}^{\prime}_{i}) {\rm exp}(z/l_i)+n_{q_{i}}(T_i),
\end{equation}
for $ q_z <0 $. Here $ l_i = | s_ {iz} | / \nu_{qi} $, where the index $ i $ takes values $ n $ or
$ f $.

Substitution of the solutions (\ref{11}) and (\ref{12}) into the boundary conditions yields a
system of four linear equations for the four coefficients $ C_{i}^{\gtrless} $. The solution of
the system is

\begin{eqnarray}\label{13}
C_{f}^{<}({\bf q'}_f)= \frac{1}{D}\Bigl\{ \alpha_1\Bigl[\Bigl(1-2\alpha_2\Bigr) - \beta_2
e^{2d_n/l_n}\Bigr][n_{q}(T_m) \,\,\,\,\,\,\,\,\,\,\,\,\nonumber \\
-n_{q}(T_B)]+ \alpha_2
e^{d_f/l_f}\Bigl(e^{2d_n/l_n}-1\Bigr)[n_{q}(T_e)-n_{q}(T_m)]\Bigr\},\,\,\,\,\,\,\,\,\,\,\,\,
\end{eqnarray}

\begin{eqnarray}\label{14}
C_{n}^{<}({\bf q'}_n)=-\frac{\alpha_2}{D}\Bigl\{ \alpha_1[n_{q}(T_m)-n_{q}(T_B)]+
\,\,\,\,\,\,\,\,\,\,\,\,\,\,\,\,\,\,\,\,\,\,\,\,\, \nonumber \\
\Bigl(e^{d_f/l_f}-\beta_1 e^{-d_f/l_f}\Bigr)[n_{q}(T_e)-n_{q}(T_m)]\Bigr\},\,\,\,\,\,\,
\end{eqnarray}

\begin{eqnarray}\label{15}
C_{f}^{>}({\bf q}_f)= \frac{e^{d_f/l_f}}{D}\Bigl\{ \beta_1
\alpha_2\Bigl(e^{2d_n/l_n}-1\Bigr)[n_{q}(T_e) -n_{q}(T_m)]  \nonumber \\
- \alpha_1
e^{d_f/l_f}\Bigl(e^{2d_n/l_n}-\beta_2\Bigr)[n_{q}(T_m)-n_{q}(T_B)]\Bigr\},
\,\,\,\,\,\,\,\,\,\,\,\,
\end{eqnarray}

\begin{eqnarray}\label{16}
C_{n}^{>}({\bf q'}_n)=-\frac{\alpha_2}{D}e^{2d_n/l_n+d_f/l_n}\Bigl\{ \alpha_1
e^{d_f/l_f}[n_{q}(T_m) \nonumber \\
-n_{q}(T_B)]+ \Bigl(e^{2d_f/l_f}-\beta_1 \Bigr)[n_{q}(T_e)-n_{q}(T_m)]\Bigr\},
\end{eqnarray}
where $ D $ denotes the determinant of the system

\begin{equation}\label{17}
D = e^{2d_n/l_n+2d_f/l_f} -\beta_1\beta_2 e^{2d_n/l_n} -\beta_2 e^{2d_f/l_f}+\beta_1(1-2\alpha_2).
\end{equation}
For compactness of expressions, we renamed $ \alpha_{f \rightarrow i} $ through $ \alpha_1 $, and
$ \alpha_{n \rightarrow f} $ through $ \alpha_2 $. In this case,  $\beta_i=1-\alpha_i$ \,\,\,\,
($i=1,2)$.

The equations for the electron temperature in the metallic layer and the magnon  temperature in
the magnetic layer are determined by the thermal balance conditions in the corresponding layers: $
P_ {ep} = W $ and $ P_ {mp} = 0 $, where $ W $ is the specific power of the heat sources in the
$N$-layer and $ P_ {ep} $ and $ P_ {mp} $ are the specific powers (averaged over the layer
thickness), that are transferred from electrons to phonons and from magnons to phonons,
respectively. $ P_ {ep} $ is expressed in terms of the phonon-electron collision integral as

 \begin{equation}\label{18}
P_{ep} = {1\over d_n} \int _{d_f<z<d_n+d_f} dz \int{{d^3 q}\over{(2\pi)^3}}\hbar\omega_{q}
I_{pe}[N_n ({\bf q}_n,z )].
\end{equation}
The substitution of (\ref{11}) and (\ref{12}) yields for the $N$-layer

\begin{eqnarray}\label{19}
W = {1\over d_n}  \int_{q_z > 0}{{d^3 q}\over{(2\pi)^3}}\hbar\omega_{nq} s_{nz}
\Bigl[C_{n}^{>}e^{-d_f/l_n} \Bigl(e^{-d_n/l_n} -1\Bigr)   \nonumber \\
-C_{n}^{<}e^{d_f/l_n}\Bigl(e^{d_n/l_n}-1\Bigr)\Bigr],\,\,\,\,\,\,\,\,\,\,\,\,
\end{eqnarray}
and for the $F$-layer we obtain

\begin{equation}\label{20}
 \int_{q_z > 0}{{d^3 q}\over{(2\pi)^3}}\hbar\omega_{fq} s_{fz} \Bigl[C_{f}^{>}
\Bigl(e^{-d_f/l_f} -1\Bigr)- C_{f}^{<}\Bigl(e^{d_f/l_f}-1\Bigr)\Bigr]=0.
\end{equation}

Taking into account the explicit form of the coefficients $ C^{\gtrless}_i $, we have

\begin{eqnarray}\label{21}
W = {1\over d_n}  \int_{q_z > 0}{{d^3 q}\over{(2\pi)^3}}\hbar\omega_{nq} s_{nz}
\Bigl(e^{d_n/l_n}-1\Bigr)\frac{\alpha_2}{D} \Bigl\{\alpha_1 e^{d_f/l_f}\,\,\,\,\,\,\,\,\,\,\,\,
\nonumber\\
\times [n_{q}(T_m)-n_{q}(T_B)]+
\Bigl(e^{2d_f/l_f}-\beta_1\Bigr)[n_{q}(T_e)-n_{q}(T_m)]\Bigr\}\,\,\,\,\,\,\,\,\,\,\,\,
\end{eqnarray}
and

\begin{eqnarray}\label{22}
\int_{q_z > 0}{{d^3 q}\over{(2\pi)^3}}\hbar\omega_{fq} s_{fz} \frac{1}{D}\Bigl(e^{d_f/l_f}-1\Bigr)
\Bigl\{\alpha_1\Bigl[ e^{2d_n/l_n} \Bigl(e^{d_f/l_f}+\beta_2\Bigr)\nonumber\\
-\beta_2 e^{d_f/l_f}-\beta_2+\alpha_2\Bigr][n_{q}(T_m)-n_{q}(T_B)]-
\,\,\,\,\,\,\,\,\,\,\,\,\nonumber\\
\alpha_2\Bigl(e^{2d_n/l_n}-1\Bigr)\Bigl(e^{d_f/l_f}+\beta_1\Bigr)[n_{q}(T_e)-n_{q}(T_m)]\Bigr\}=0.
\,\,\,\,\,\,\,\,\,\,\,\,
\end{eqnarray}

In the case of thick $F$ and $N$ layers, when $ d_f \gg l_f $ and $ d_n \gg l_n $, Eq.~(\ref{21})
is greatly simplified and  reduces to

\begin{eqnarray}\label{23}
W = {\frac{1}{d_n}}  \int_{0}^{\infty} {{q^2 dq}\over{(2\pi)^2}}\int_{0}^{\pi/2}\sin\theta d\theta
\cos\theta\alpha_2 (\theta) s_n \hbar\omega_{q} \nonumber\\ \times
[n_{q}(T_e)-n_{q}(T_m)].\,\,\,\,\,\,\,\,\,\,\,\,
\end{eqnarray}

If we introduce the probability $\alpha_2$ averaged over the angles of incidence

\begin{equation}\label{23}
\langle\alpha_2\rangle = 2  \int_{0}^{\pi/2}\sin\theta d\theta \cos\theta\alpha_2 (\theta),
\end{equation}
then integration over the phonon wave vectors gives

\begin{equation}\label{24}
Wd_n \equiv Q = \frac{\pi^2\langle\alpha_2\rangle}{120\hbar^3s_{n}^{2}}
\Bigl(T_{e}^{4}-T_{m}^{4}\Bigr).
\end{equation}

For weak heating, when $ T_{e} -T_{B} \ll T_{B} $ and $ T_{m} -T_{B} \ll T_{B} $,  from
Eq.~(\ref{24}) we obtain the temperature difference on the $ N / F $-boundary as

\begin{equation}\label{25}
T_{e}-T_{m} = \frac{30\hbar^3 s_{n}^{2}}{\pi^2\langle\alpha_2\rangle}\frac{W d_n }{T_{B}^{3}}.
\end{equation}

We now consider the case of thin layers, when $ d_f \ll l_f $ and $ d_n \ll l_n $. In this limit,
the determinant $ D = \alpha_1 \alpha_2 $, and this leads to the following equation for the
electron temperature:

\begin{equation}\label{26}
W =   \int_{q_z > 0}{{d^3 q}\over{(2\pi)^3}}\hbar\omega_{qn} \nu_{qn}
 [n_{q}(T_e)-n_{q}(T_B)].
\end{equation}
Eq.~(\ref{26}) exactly coincides with the  equation for the electron temperature in a thin metal
film lying on a dielectric substrate (see, for example, \cite{Bezuglyj-Shklovskij-JETF1997}). The
integral over the phonon wave vectors  gives

\begin{equation}\label{27}
W = \Sigma_5 (T_{e}^{5}-T_{B}^{5}) ,
\end{equation}
where the electron-phonon coupling constant is

\begin{equation}\label{28}
 \Sigma_5 =  {\frac{D_5 m^2 \mu^2  }{4\pi^3 \rho  \hbar^7 s^4}}.
\end{equation}
The number $ D_5 \approx 24.9 $ is the integral $ D_k = \int_{0}^{\infty} x^{k-1} (e^x-1)^{-1} \,
dx $ for $ k = 5 $. As can be seen from (\ref{27}), a thin layer of a ferromagnet has no effect on
the kinetics of nonequilibrium phonons, since they are almost not absorbed in it.

The magnon temperature $ T_m $ in the $F$ layer is determined by  Eq.~(\ref{22}). It turns out
that in the case of thin layers $ T_m = T_B $, i.e. the magnon temperature coincides with the
temperature of the thermostat. The reason for this is clear: the nonequilibrium phonons emitted in
the metallic layer go into the substrate without being absorbed by the magnons in the ferromagnet
layer, and therefore, do not heat the magnon gas.

Since in the case of thin layers $ T_m = T_B $, the temperature difference between electrons and
magnons for thin layers can be obtained from Eq.~(\ref{27}). At low temperatures, we have $ T_ {e}
-T_ {m} = {W} / {5 \Sigma_5 T_{B}^{4}} $, from which it follows that the temperature jump on the $
N / F $ boundary increases as the temperature of the thermostat $ T_B $ decreases. Since the
voltage on a normal metal under the conditions of both the longitudinal and transverse spin
Seebeck effects is proportional to the difference $ T_ {e} -T_ {m} $
\cite{Schreier-Kamra-PhysRev2013, Xiao2010}, this voltage should increase as the temperature of
the thermostat is lowered.

Experimental studies of the spin Seebeck effect are often realized on $ N / F $ heterostructures
in which the $ F $-layer has a rather large thickness, whereas the $ N $-layer is thin.  To apply
our results to this case, we consider the limit $ d_f \gg l_f $ and $ d_n \ll l_n $. Then, the
magnon temperature is determined by equation

\begin{equation}\label{29}
Wd_n = \frac{\pi^2\langle\alpha_1\rangle}{120\hbar^3s_{f}^{2}}\Bigl(T_{m}^{4}-T_{B}^{4}\Bigr)
\equiv\Sigma_4 (T_{m}^{4} -T_{B}^{4}),
\end{equation}
and for the electron temperature we have

\begin{equation}\label{30}
W = \Sigma_5 (T_{e}^{5}-T_{m}^{5}).
\end{equation}
As in the case of thin layers, the temperature jump on the $ N / F $-boundary for  weak heating is
given by $ T_{e} -T_{m} = {W} / {5 \Sigma T_{B}^{4}} $, i.e. this jump rapidly increases as the
temperature of the thermostat $ T_B $ decreases.

\section{Heat transfer through a two-layer N/F system located between two insulators with different
temperatures}\label{s2}

\begin{figure}[t]
  \includegraphics[width=9.5 cm]{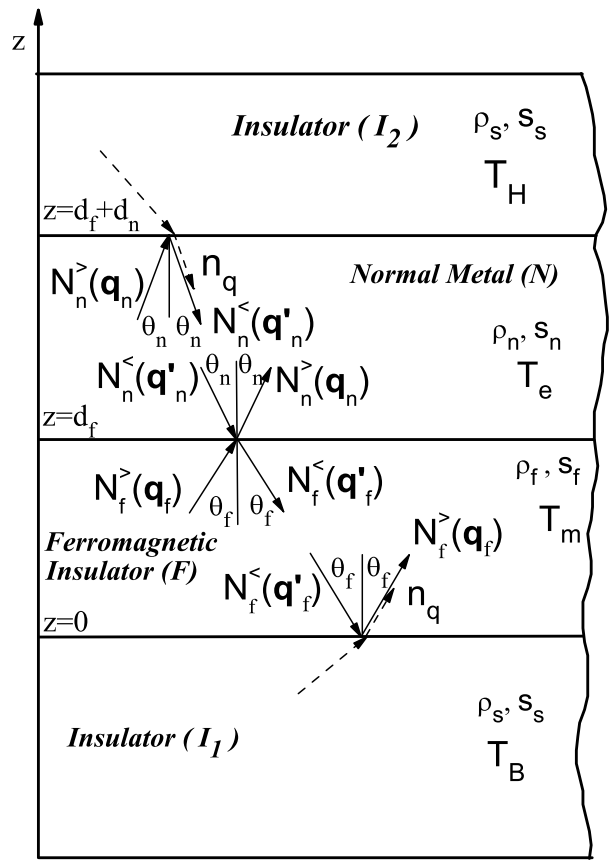}
  \caption{Refraction and reflection of phonon modes at interfaces in a layered $I_1 / N / F / I_2$
structure. $ I_1 $ is a massive dielectric substrate (heat sink with temperature $ T_B $), $ F $
is a ferromagnetic insulator, $ N $ is a normal metal, $ I_2 $ is a heat source in the form of a
thick dielectric layer with temperature $ T_H> T_B $. The remaining notations are the same as in
Fig.~\ref{Fig1}.} \label{Fig2}
\end{figure}

Now we consider the layered structure $ I_1 / N / F / I_2 $ in which a metal layer with a
thickness of $ d_n $ and a layer of a ferromagnetic insulator with a thickness $ d_f $ are placed
between two massive dielectrics $ I_1 $ and $ I_2 $. The temperatures of the dielectrics  are $
T_B $ and $ T_H $ with $ T_H> T_B $ (see Fig.~\ref{Fig2}).

As in Section~\ref{s1}, we assume that the electrons in the $ N $-layer and the magnons in the $ F
$-layer are thermalized and have temperatures $ T_e $ and $ T_m $, respectively. Again, we will
assume that, owing to the high electron and magnon thermal conductivities, the temperatures $ T_e
$ and $ T_m $ do not depend on the coordinate $ z $ perpendicular to the
layers.\cite{Akhieser_Shishkin} In contrast, the phonon distribution functions in the $N$-layer
and in the $F$-layer depend on $ z $, and this dependence is determined by the corresponding
kinetic equations. For the $N$-layer, the kinetic equation has form (\ref{1}), and for the
$F$-layer it has form (\ref{1aa}). Each of these equations can be sub-divided into two equations
by writing them for $ N^{>} ({\bf q}, z) $ and $ N^{<} ({\bf q '}, z) $,  that is for the
distribution functions of the phonons with positive and negative projections of the wave vector on
the $ z $ axis. To find the phonon distribution functions, the boundary conditions must be added
to Eqs.~(\ref{1}) and (\ref{1aa}). For the boundaries $ z = 0 $ and $ z = d_f $, they do not
differ from the expressions (\ref{4}), (\ref{6}) and (\ref{7}) given in Sec.~\ref{s1}:

\begin{equation}\label{201}
N^{>}_{f} ({\bf q}_f, 0)=\alpha_{1} n_{q}(T_B)+\beta_{1} N^{<}_{f} ({\bf q'}_f, 0),
\end{equation}

\begin{equation}\label{202}
N^{<}_{f} ({\bf q'}_f, d_f)=\alpha_{2} N^{<}_{n} ({\bf q'}_n, d_f)+\beta_{2} N^{>}_{f} ({\bf q}_f,
d_f),
\end{equation}

\begin{equation}\label{203}
 N^{>}_{n} ({\bf q}_n, d_f)=\alpha_{2} N^{>}_{f} ({\bf q'}_f,
d_f)+\beta_{2} N^{<}_{n} ({\bf q}_n, d_f).
\end{equation}
For $ z = d_f + d_n $, the boundary condition has the same structure as for $ z = 0 $:

\begin{equation}\label{204}
N^{<}_{n} ({\bf q'}_n, d_f+d_n)=\beta_3 N^{>}_{n} ({\bf q}_n, d_f+d_n)+\alpha_3 n_q(T_H).
\end{equation}

Substitution of the solutions (\ref{11}), (\ref{12}) into the boundary conditions (\ref{201}),
(\ref{202}), (\ref{203}) and (\ref{204} ) gives the following expressions for the coefficients $
C^{\gtrless} $:

\begin{eqnarray}\label{205}
C_{f}^{<}({\bf q'}_f)= \frac{1}{D}\Bigl\{ \alpha_2\alpha_3 e^{d_n/l_n+d_f/l_f}
[n_{q}(T_H)-n_{q}(T_e)]+ \nonumber \\
\alpha_2 e^{d_f/l_f}\Bigl(e^{2d_n/l_n}-\beta_3\Bigr)[n_{q}(T_e)-n_{q}(T_m)] -\alpha_1
\Bigl[\beta_2e^{2d_n/l_n} \nonumber
\\-(\beta_2-\alpha_2)\beta_3\Bigr][n_{q}(T_m)-n_{q}(T_B)]\Bigr\},\,\,\,\,\,\,\,\,\,\,\,\,\,\,\,\,
\end{eqnarray}

\begin{eqnarray}\label{206}
C_{n}^{<}({\bf q'}_n)=\frac{e^{-d_f/l_f}}{D}\Bigl\{ \alpha_3
e^{d_n/l_n+d_f/l_f}\Bigl(e^{d_f/l_f}-\beta_1\beta_2e^{-d_f/l_f}\Bigr) \nonumber
\\ \times[n_{q}(T_H)-n_{q}(T_e)]
-\alpha_2\beta_3\Bigl(e^{2d_f/l_f}-\beta_1 \Bigr)[n_{q}(T_e)\nonumber
\\-n_{q}(T_m)]-\alpha_1\alpha_2\beta_3e^{d_f/l_f}[n_{q}(T_m)-n_{q}(T_B)]\Bigr\},
\,\,\,\,\,\,\,\,\,\,\,\,\,\,\,\,
\end{eqnarray}

\begin{eqnarray}\label{207}
C_{f}^{>}({\bf q}_f)=
\frac{e^{d_f/l_f}}{D}\Bigl\{\beta_1\alpha_2\alpha_3e^{d_n/l_n}[n_{q}(T_H)-n_{q}(T_e)]\nonumber \\
+\beta_1\alpha_2 \Bigl(e^{2d_n/l_n}-\beta_3\Bigr) [n_{q}(T_e)-n_{q}(T_m)]\nonumber
\\-\alpha_1e^{d_f/l_f}\Bigl(e^{2d_n/l_n}-\beta_2\beta_3\Bigr) [n_{q}(T_m)-n_{q}(T_B)]\Bigr\},
\,\,\,\,\,\,\,\,\,
\end{eqnarray}

\begin{eqnarray}\label{208}
C_{n}^{<}({\bf
q'}_n)=\frac{e^{d_n/l_n+d_f/l_n}}{D}\Bigl\{\alpha_3\Bigl[e^{2d_f/l_f}-\beta_1(\beta_2-\alpha_2)\Bigr]
\nonumber \\
\times[n_{q}(T_H)-n_{q}(T_e)] -\alpha_2 e^{d_n/l_n}\Bigl(e^{2d_f/l_f}-\beta_1
\Bigr)[n_{q}(T_e)\nonumber \\-n_{q}(T_m)]-\alpha_1 \alpha_2
e^{d_n/l_n+d_f/l_n}[n_{q}(T_m)-n_{q}(T_B)] \Bigr\},\,\,\,\,\,\,\,\,\,
\end{eqnarray}
where $ D $ denotes the determinant of the system

\begin{eqnarray}\label{209}
D = e^{2d_n/l_n+2d_f/l_f} -\beta_1\beta_2 e^{2d_n/l_n} -\beta_2\beta_3 e^{2d_f/l_f}\nonumber
\\+\beta_1(1-2\alpha_2)\beta_3.\,\,\,\,\,\,\,\,\,
\end{eqnarray}

To find the temperatures $ T_e $ and $ T_m $, we use the condition that on all interfaces the heat
flux is the same. If we denote the $ z $-component of the heat flux as $ Q_z (z) $, then the
equations for $ T_e $ and $ T_m $ can be written as

\begin{equation}\label{210}
Q_z(d_f - 0) = Q_z(+0),
\end{equation}

\begin{equation}\label{211}
Q_z(s=d_f +d_n - 0) = Q_z(d_f +0).
\end{equation}
The $ z $-component of the phonon heat flux in the $ N $-layer is expressed in terms of the phonon
distribution function as

\begin{equation}\label{212}
Q_z = \int_{q_z>0}{{d^3 q}\over{(2\pi)^3}}\hbar\omega_q s_{nz}[N^{>}_{n} ({\bf q}_n,z)-N^{<}_{n}
({\bf q'}_n,z)]
\end{equation}
The expression for the phonon heat flux in the $ F $-layer has a completely similar form. Note
that the equality $ Q_z (d_f -0) = Q_z (d_f +0) $ is automatically satisfied (see
\cite{Little1959}).

Substitution of the phonon distribution functions into Eq.~(\ref{210}) gives

\begin{eqnarray}\label{213}
 \int_{q_z>0}{{d^3 q}\over{(2\pi)^3}}\hbar\omega_q s_{fz}\frac{\Bigl(e^{d_f/l_f}-1 \Bigr)}{D}
 \Bigl\{\alpha_2\alpha_3e^{d_n/l_n}\Bigl(e^{d_f/l_f}+\beta_1\Bigr)\nonumber \\
\times [n_{q}(T_H)-n_{q}(T_e)] +\alpha_2 \Bigl(e^{2d_n/l_n}-\beta_3\Bigr)
\Bigl(e^{d_f/l_f}+\beta_1\Bigr)\nonumber \\ \times[n_{q}(T_e)-n_{q}(T_m)]
-\alpha_1\Bigl(e^{2d_n/l_n+d_f/l_f}-\beta_2\beta_3 e^{d_f/l_f}\nonumber \\
+ \beta_2e^{2d_n/l_n}-(\beta_2 -
\alpha_2)\beta_3\Bigr)[n_{q}(T_m)-n_{q}(T_B)]\Bigr\}=0.\,\,\,\,\,\,\,\,\,\,\,\,\,\,\,\,\,\,
\end{eqnarray}

After a similar substitution of the corresponding phonon distribution functions into the
Eq.~(\ref{211}), we obtain

\begin{eqnarray}\label{214}
 \int_{q_z>0}{{d^3 q}\over{(2\pi)^3}}\hbar\omega_q s_{nz}\frac{\Bigl(e^{d_n/l_n}-1 \Bigr)}{D}
 \Bigl\{\alpha_1\alpha_2e^{d_f/l_f}\Bigl(e^{d_n/l_n}+\beta_3\Bigr)\nonumber \\
 \times[n_{q}(T_m)-n_{q}(T_B)]
+\alpha_2 \Bigl(e^{2d_f/l_f}-\beta_1\Bigr) \Bigl(e^{d_n/l_n}+\beta_3\Bigr) \nonumber \\
\times [n_{q}(T_e)-n_{q}(T_m)] -\alpha_3\Bigl(e^{2d_f/l_f+d_n/l_n}-\beta_1\beta_2 e^{d_n/l_n}
\nonumber \\
+ \beta_2e^{2d_f/l_f}-\beta_1(\beta_2 - \alpha_2)\Bigr)[n_{q}(T_H)-n_{q}(T_e)]\Bigr\}=0.
\,\,\,\,\,\,\,\,\,\,\,\,\,\,\,\,\,\,
\end{eqnarray}

In the case of thick layers ($ d_n \gg l_n $, $ d_f \gg l_f $), the equations for $ T_e $ and $
T_m $ have the form
\begin{eqnarray}\label{215}
\int_{q_z>0}{{d^3 q}\over{(2\pi)^3}}\hbar\omega_q s_{fz}\{\alpha_2[n_{q}(T_e)-n_{q}(T_m)]\nonumber
\\-\alpha_1[n_{q}(T_m)-n_{q}(T_B)]\}=0,
\end{eqnarray}

\begin{eqnarray}\label{216}
\int_{q_z>0}{{d^3 q}\over{(2\pi)^3}}\hbar\omega_q s_{nz}\{\alpha_3[n_{q}(T_H)-n_{q}(T_e)]\nonumber
\\-\alpha_2[n_{q}(T_e)-n_{q}(T_m)]\}=0.
\end{eqnarray}
Integration over the phonon wave vectors gives

\begin{equation}\label{217}
\langle\alpha_2\rangle_f(T_{e}^4 - T_{m}^4)=\langle\alpha_1\rangle_f(T_{m}^4 - T_{B}^4),
\end{equation}

\begin{equation}\label{218}
\langle\alpha_3\rangle_n(T_{H}^4 - T_{e}^4)=\langle\alpha_2\rangle_n(T_{e}^4 - T_{m}^4).
\end{equation}
As before, the notations $ \langle \alpha \rangle_f $ and $ \langle \alpha \rangle_n $ denote  the
averaging over the angles of phonon incidence $ \theta_f $ or $ \theta_n $ of the probability of
the phonon passing through the corresponding interlayer boundary. For example,

\begin{equation}\label{219}
\langle\alpha_2\rangle_n=2\int_{0}^{\pi/2}\sin (\theta_n)\,\cos
(\theta_n)\,\,\alpha_2(\theta_n)\,\,d\theta_n.
\end{equation}
Using Snell's law $ s_n \sin (\theta_f) = s_f \sin (\theta_n) $, one can show that
$ s_{f}^2 \langle \alpha_2 \rangle_n = s_{n}^2 \langle \alpha_2 \rangle_f $.

The solutions of the set of Eqs.~(\ref{217}), (\ref{218}) are
\begin{equation}\label{220}
T^{4}_m=\frac{\langle\alpha_2\rangle_f \langle\alpha_3\rangle_n T^{4}_H
+\langle\alpha_1\rangle_f(\langle\alpha_2\rangle_n +\langle\alpha_3\rangle_n)T^{4}_B}
{\langle\alpha_1\rangle_f\langle\alpha_2\rangle_n+
\langle\alpha_1\rangle_f\langle\alpha_3\rangle_n +\langle\alpha_2\rangle_f
\langle\alpha_3\rangle_n},
\end{equation}

\begin{equation}\label{221}
T^{4}_e=\frac{(\langle\alpha_1\rangle_f+\langle\alpha_2\rangle_f) \langle\alpha_3\rangle_n T^{4}_H
+\langle\alpha_1\rangle_f\langle\alpha_2\rangle_n T^{4}_B}
{\langle\alpha_1\rangle_f\langle\alpha_2\rangle_n+
\langle\alpha_1\rangle_f\langle\alpha_3\rangle_n +\langle\alpha_2\rangle_f
\langle\alpha_3\rangle_n}.
\end{equation}

In the case when the temperatures $ T_{H} $ and $ T_{B} $ differ only slightly, so that $
T_{H} -T_{B} \ll T_{B} $, the temperature difference between electrons and magnons is given by

\begin{equation}\label{222}
T_e-T_m=\frac{\langle\alpha_1\rangle_f \langle\alpha_3\rangle_n (T_H-T_B)}
{\langle\alpha_1\rangle_f\langle\alpha_2\rangle_n+
\langle\alpha_1\rangle_f\langle\alpha_3\rangle_n +\langle\alpha_2\rangle_f
\langle\alpha_3\rangle_n}.
\end{equation}

We now turn to the limiting case of thin layers, when $ d_f \ll l_f $ and $ d_n \ll l_n $. In this
limit, the equations for $ T_{e} $ and $ T_{m} $ have the following form:
\begin{eqnarray}\label{223}
\int_{q_z > 0}{{d^3 q}\over{(2\pi)^3}}\hbar\omega_{qn}
\nu_{qn}\frac{1}{D}\Bigl\{\alpha_1\alpha_2(1+\beta_3)[n_{q}(T_e)-n_{q}(T_B)] \nonumber \\
-\alpha_3(1+\beta_1\alpha_2 +\beta_2\alpha_1-\beta_1\beta_2)[n_{q}(T_H)-n_{q}(T_e)]\Bigr\} = 0,
\,\,\,\,\,\,\,\,\,\,\,\,
\end{eqnarray}

\begin{eqnarray}\label{224}
\int_{q_z > 0}{{d^3 q}\over{(2\pi)^3}}\hbar\omega_{qf}
\nu_{qf}\frac{1}{D}\Bigl\{\alpha_2\alpha_3(1+\beta_1)[n_{q}(T_H)-n_{q}(T_m)] \nonumber \\
-\alpha_1(1+\beta_2\alpha_3 +\beta_3\alpha_2-\beta_2\beta_3)[n_{q}(T_m)-n_{q}(T_B)]\Bigr\} = 0,
\,\,\,\,\,\,\,\,\,\,\,\,
\end{eqnarray}
where $D = 1-\beta_1\beta_2-\beta_2\beta_3+\beta_1(\beta_2-\alpha_2)\beta_3$.

It is interesting that, unlike the case of continuous heating of the $N$-layer considered in
Sec.~\ref{s1}, in the $ I_1 / N / F / I_2 $ heterostructure with two massive dielectric plates
maintained at different temperatures, the electron and magnon temperatures depend on the acoustic
transparencies of  interfaces even for arbitrarily small layer thicknesses $ d_f $ and $ d_n $.
This can be easily seen by noting that the layer thicknesses  $ d_f $ and $ d_n $ drop out of
Eqs.~(\ref{223}) and (\ref{224}).

In the integrals (\ref{223}) and (\ref{224}), the integration over angles is separated from the
integration over the absolute value of the phonon wave vector. Because of the  simple linear
relationship between $ \nu_{qn} $ and $ \omega_{qn} $, the equation for the electron temperature
reduces to the following form:

\begin{equation}\label{225}
T^{5}_e=\frac{\Psi_{31} T^{5}_H +\Phi_{13} T^{5}_B} {\Psi_{31}+ \Phi_{13}},
\end{equation}
where we have denoted

\begin{equation}\label{226}
\Phi_{13}=\int_{0}^{\pi/2}d\theta\,\sin(\theta)\alpha_1\alpha_2(1+\beta_3)/D,
\end{equation}
and

\begin{equation}\label{227}
\Psi_{31}=\int_{0}^{\pi/2}d\theta\,\sin(\theta)\alpha_3(1+\beta_1\alpha_2
+\beta_2\alpha_1-\beta_1\beta_2)/D.
\end{equation}

The equation for the magnon temperature has a more complicated form, namely:

\begin{equation}\label{228}
F(T_m)=\frac{\Phi_{31} F(T_H) +\Psi_{13} F(T_B)} {\Phi_{31}+ \Psi_{13}}.
\end{equation}
Here, indices $ 1 \leftrightarrow 3 $ are swapped in $ \Phi $  and $ \Psi $, and $ F (T) $ is a
function of temperature that is determined by the integral

\begin{equation}\label{229}
F(T)=\int_{0}^{\infty}q^3 \nu_{qf}(T) n_q(T)\,dq.
\end{equation}
In the case when the heating of the system is weak, i.e. $ T_H - T_B \ll T_B $, the expression
for the temperature difference between electrons and magnons is substantially simplified and can
be written as

\begin{equation}\label{230}
T_e - T_m =\Bigl (\frac{\Psi_{31}}  {\Psi_{31}+ \Phi_{13}}- \frac{\Phi_{31}}  {\Phi_{31}+
\Psi_{13}}\Bigr )(T_H - T_B).
\end{equation}

\begin{figure}[t]
  \includegraphics[width=9.5 cm]{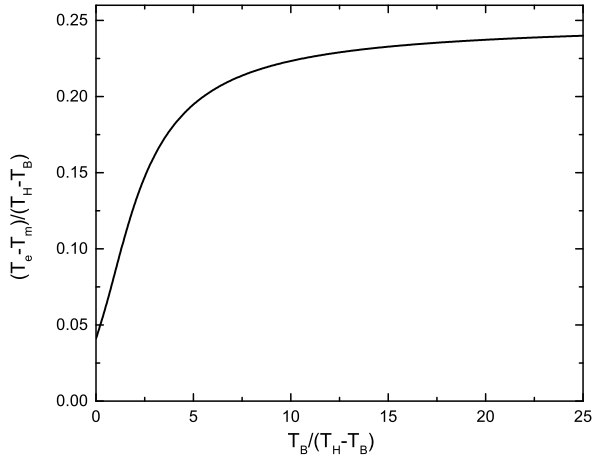}
\caption{The dependence of the difference between the electron and magnon temperatures on the
thermostat temperature for thin $F$- and $N$-layers. In order not to complicate the physical
picture, we put $ \alpha_1 (\theta) = \alpha_2 (\theta) = \alpha_3 (\theta) = \alpha (\theta $).
For $ \alpha (\theta) $, the stepwise approximation is used: $ \alpha (\theta) = 1/2 $ for $
\theta <\theta_{cr} $ and $ \alpha (\theta) = 0 $ for $ \theta> \theta_{cr} $ (where $
\theta_{cr}$ is the angle of total internal reflection).~\cite{Bezuglyj-Shklovskij-JPhys2018} At
large $T_B/(T_H - T_B)$, the curve approaches asymptotically to~0.25.  }\label{Fig3a}
\end{figure}

Dependence of $ (T_e -T_m) / (T_H -T_B) $ on $ T_B) / (T_H -T_B) $ is shown in the Fig.
\ref{Fig3a}. It is seen that for $ T_B \gg T_H - T_B $ this dependence goes to a constant value,
which  is described by the Eq.~(\ref{230}).

As in the previous section, we consider the case of $ d_f \gg l_f $, $ d_n \ll l_n $, which is
important for the experiment. In this case, the determinant $ D = e^{2d_f / l_f} (1- \beta_2
\beta_3) $, and the equations for the magnon and electron temperatures have the form

\begin{eqnarray}\label{231}
 \int_{q_z>0}{{d^3 q}\over{(2\pi)^3}}\hbar\omega_q s_{fz}\{\alpha_1
 [n_{q}(T_m)-n_{q}(T_B)]\nonumber \\-\frac{\alpha_2 \alpha_3}{(1-\beta_2\beta_3)}
 [n_{q}(T_H)-n_{q}(T_m)]\}=0,
\end{eqnarray}

\begin{eqnarray}\label{232}
 \int_{q_z>0}{{d^3 q}\over{(2\pi)^3}}\frac{\hbar\omega_q \nu_{nq}}{(1-\beta_2\beta_3)}
 \{\alpha_2(1+\beta_3)
 [n_{q}(T_e)-n_{q}(T_m)]\nonumber \\-\alpha_3(1+\beta_2) [n_{q}(T_H)-n_{q}(T_e)]\}=0.
 \,\,\,\,\,\,\,\,\,\,\,\,
\end{eqnarray}
In both equations, the integration over the angles of the phonon wave vector is separated from the
integration over its absolute value. As a result of the integration,  we obtain  for the magnon
temperature

\begin{equation}\label{233}
 \langle{\alpha_1}\rangle
 (T_{m}^4-T_{B}^4)=\langle{\frac{\alpha_2 \alpha_3}{1-\beta_2\beta_3}}\rangle(T_{H}^4-T_{m}^4),
\end{equation}
and the Eq.~(\ref{232}) gives

\begin{equation}\label{233}
T_{e}^5=\frac{1}{2}\Bigl [ T_{m}^5\,\overline{\frac{\alpha_2
(1+\beta_3)}{1-\beta_2\beta_3}}+T_{H}^5\,\overline{\frac{\alpha_2
(1+\beta_3)}{1-\beta_2\beta_3}}\Bigr ],
\end{equation}
where we have denoted
$$\overline{f(\theta)}=\int_{0}^{\pi /2}\sin(\theta)f(\theta)\,d\theta.$$

For weak heating, when $ T_{H} -T_{B} \ll T_{B} $, the difference between the electron and magnon
temperatures is given by

\begin{equation}\label{234}
T_{e}-T_{m}=\frac{1}{2}\,\overline{\frac{\alpha_3
(1+\beta_2)}{1-\beta_2\beta_3}}\,\frac{\langle\alpha_1\rangle(T_{H}-T_{B})}{\langle\alpha_1\rangle
+ \langle{\alpha_2 \alpha_3}/{(1-\beta_2\beta_3)}\rangle}\,.
\end{equation}

Thus, we have obtained expressions for the electron and magnon temperatures for two different sets
of experiments: (1) when in a heterostructure  "normal metal-ferromagnetic insulator-dielectric
substrate", the metal layer is heated by a direct current \cite{Schreier-APL2013, Wang-APL2014},
and (2) when the "normal metal-ferromagnetic dielectric" bilayer is enclosed between two massive
dielectric plates that are maintained at different temperatures.

\section{Main conclusions and comparison with experiment }\label{s3}

When comparing the results with experiment, one should keep in mind that there are two different
mechanisms for generating a spin current by a heat flux. The first of them is connected with the
difference between the electron and magnon temperatures on the $ N / F $-interface. The second
mechanism stems from   the magnon temperature gradient in the ferromagnetic insulator. At low
temperatures $T_B \ll \Theta_D$ the first mechanism will dominate due to the two reasons: a) the
Kapitza thermal resistance of the $N/F$-boundary increases as $ T_ {B} ^ {-3} $ and  b) the
thermal conductivity of the $F$-plate strongly increases \cite{Akhieser_Shishkin}. The latter
reason allows us to consider $T_m = const$, i.e. to neglect the gradient of $T_m$ at low
temperatures. In this context, we would like to quote the  results of Ref. \cite{Kimling2017},
where  LSSE measurements on a picosecond time scale in the temperature range from 295K to 473K
have been reported. In Ref. \cite{Kimling2017} it was shown that the spin current $J_S \approx
\alpha_S (T_e - T_m)$, where  $\alpha_S (T)$ monotonically decreases with the temperature and
vanishes approximately at Curie temperature  $\Theta_C$ for YIG, as expected from the temperature
dependence of  Kapitza interfacial resistance. Also, from the independence of $\alpha_S$ of
thickness of the magnetic insulator (for 20 to 250 nm thick YIG samples), it follows that the
contribution from the bulk LSSE is negligible on picosecond time scales. So, in
Ref.~\cite{Kimling2017}  the interfacial mechanism of LSSE  has been separated from bulk LSSE
contribution with $\nabla T_m \neq 0$. As the generation of the spin current by a temperature
difference $T_e - T_m$ can arise due to heating of the $N$-layer by an ultrashort (femtosecond)
laser pulse \cite{Kimling2017}, the reason for this generation is that at such  short times the
gradient of the magnon temperature does not have time to form. The model presented here can be
used to describe non-stationary experiments (see, for example, Ref.~\cite{Bezuglyj2018}).

Within the framework of the considered model, the only mechanism for creating a spin current is
the temperature difference between magnons and electrons on the $ F / N $ interface. Thus, if $
T_e \neq T_m $, the magnon gas is not in equilibrium with the electron gas and the magnon
absorption by electrons is not compensated by their emission. As a consequence, the interaction of
nonequilibrium magnons with electrons leads to spin polarization of the electron gas near the $ F
/ N $ interface and to the subsequent diffusion of spin-polarized electrons into the $ N $-layer,
i.e. to the spin current $ {\bf J} _s $. As a result of the spin-orbit interaction, the spin
current generates a charge current in the perpendicular direction and a potential difference $ V_
{ISHE} $ emerges at the lateral edges of the $ N $-layer. When the electron and magnon temperature
difference $ \Delta T $ is small in comparison with the substrate temperature, the value $
V_{ISHE} \propto \Delta T $. From the results of Section \ref{s1} it follows that for a continuous
heating of the $ N $-layer, the difference between the electron and magnon temperatures is
proportional to the specific heating power $ W $, and hence $V_{ISHE} \propto W$. This linear
dependence was observed in the experiment \cite{Schreier-APL2013, Wang-APL2014}. In the case when
the $ F / N $ system is enclosed between two heat reservoirs with temperatures $ T_H $ and $ T_B
$, and $ T_H > T_B $, the value $ V_{ISHE} \propto (T_H-T_B) $ (see the experimental works
\cite{Uchida2010-1, Kikkawa2013, Rezende2014}). According to the results of Section~\ref{s2}, the
temperature difference $ T_e - T_m $ is proportional to $ T_H-T_B $ (see Eq.~(\ref{230})). Thus,
in the case of the $ I_1 / F / N/I_2 $ sandwich, the results of our theory agree with the
experiment. Also, the dependence $ T_e -T_m $ on $ T_B $, presented in Fig. \ref{Fig3a}, agrees
with the previous calculations.\cite{Shen2016}

We note an interesting feature that is predicted in  Section \ref{s1}, namely, the growth of $ T_e
- T_m $ (and therefore, also $ V_{ISHE} $) when the temperature of the thermostat $ T_B $
decreases at the fixed heating power $ W $. The reason for this growth is an increase in the
thermal resistance of the interfaces when the temperature is lowered. This behavior of $ V_{ISHE}
(T_B) $ was observed in the experiments \cite{Guo-Cramer_PhysRev2016}.

\begin{figure}[t]
\includegraphics[width=9.5 cm]{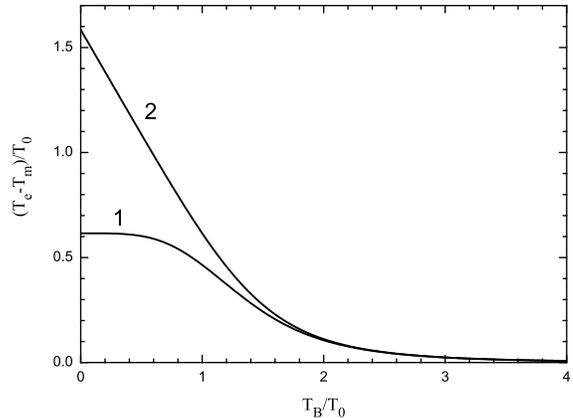}
\caption{The dimensionless difference between the electron and magnon temperatures  for thick
$F$-layers (1) and for thin $F$-layers (2).} \label{Fig3}
\end{figure}

Since we believe that the magnon temperature is uniform in the thickness of the $ F $-layer, the
results obtained should be compared with experiments performed at low temperatures. Figure
\ref{Fig3} shows the dependence of the temperature jump $ T_e - T_m $ on the temperature of the
thermostat. In the case of thick $ F $ -layers, when $ d_f \gg l_f $ (but $ d_n \ll l_n $), the
temperature difference between electrons and magnons is determined by Eq. (\ref{30}), and the
temperature difference between magnons and thermostat is given by Eq. (\ref{29}). From these
equations it follows that

\begin{equation}\label{234}
 \frac{T_e - T_m }{T_0} =[w+(1+t_B)^{5/4}]^{1/5}-(1+t_{B}^{4})^{1/4}.
\end{equation}
Here, the dimensionless temperature of the thermostat is $ t_B = T_B / T_0 $; $ T_0 = (W
/\Sigma_4)^{1/4} $. The notation $ w = \Sigma_{4}^{5/4} / \Sigma_5 W $ is introduced.

In the case when both layers are thin, i.e. $ d_n \ll l_n $ and $ d_f \ll l_f $, the
dimensionless difference between the electron and magnon temperatures has the form

\begin{equation}\label{235}
 \frac{T_e - T_m }{T_0} =(w+t_{B}^{5})^{1/5} - t_B.
\end{equation}
The dependencies (\ref{234}) and (\ref{235}) are shown in Fig. \ref{Fig3}.

In experiments on the spin Seebeck effect, the voltage $ V_{ISHE} $  in the $N$-layer, which
generates a spin current due to the inverse spin Hall effect (ISHE), is usually measured. Since
the magnitude of the spin current through the $ N / F $-boundary is proportional to the
temperature difference $ T_e -T_m $,\cite{Xiao2010, Kimling2017, Adachi2013, Bender2015} and the
voltage $ V_{ISHE} $ is proportional to the spin current \cite{Saitoh2006}, then $ V_{ISHE} = K
(T_e -T_m) $, where the coefficient proportionality $ K $ depends on the temperature of the
thermostat. The direct proportionality between $ V_ {ISHE} $ and $ (T_e -T_m) $ allows us to
compare our results with experiment. According to Fig. \ref{Fig3} at the same temperature of the
thermostat $ T_B $, the difference between the electron and magnon temperatures $ (T_e -T_m) $ is
smaller for thicker $ F $-layers. This result agrees with the experiment \cite{Prakash2018}, where
a decrease in the electric field strength in the $ N $-layer was observed when the thickness of
the $ F $-layer increased from 100 nm to 1 $\mu$m. Also, in the experiment from
Ref.~\cite{Prakash2018}, $ V_ {ISHE} $ growth was observed for thin $ F $-layers at $ T_B
\rightarrow 0 $. This behavior of $ V_ {ISHE} $ is consistent with the dependence $ (T_e -T_m) $
on $ T_B $, represented by curve 2 in Fig. \ref{Fig3}.

As can be seen from Fig. \ref{Fig3} here and from Fig. 1 in Ref.~\cite{Prakash2018}, the
discrepancy between the results of the theory and experiment takes place at high temperatures $
T_B $, when in the theory $ (T_e -T_m) \rightarrow 0 $, and in the experiment $ V_{ISHE}
\rightarrow const $. In our opinion, this difference is due to the fact that for large $ T_B $,
the spin current is mainly generated by the temperature gradient in the $ F $-layer. Since the
temperature gradient is not taken into account in our theory its range of applicability is limited
to the low-temperature region. So, at low $T_B$, curve (1) agrees with experiment on the bulk
sample, and curve (2) agrees with experiments on thin layers of YIG. The difference between the
experimental and theoretical dependences at high $T_B$ is due to the dominance of the magnon
temperature gradient over the $T_e-T_m$ temperature difference in the generation of the spin
current. In the case of heating of the N-layer, the dependence of $T_e-T_m$ on the temperature of
the thermostat (Fig. \ref{Fig3})  differs qualitatively from the same dependence in the case of a
fixed temperature difference $T_H-T_B$ (Fig.~\ref{Fig3a}). This physical picture is consistent
with the experimental results presented in Fig.~1 in Ref.~\cite{Prakash2018} and in Fig.~6 in
Ref.~\cite{Rezende2014}.

Finally, let us briefly  discuss another aspect of applicability of our model, which operates with
concepts of electron and magnon temperatures. As noticed earlier
\cite{Bezuglyj-Shklovskij-JETF1997, Bezuglyj-ShklovskijFNT2016}, the electron temperature can be
introduced if the thermalization time of electrons, i.e. the time of electron-electron collisions
$ \tau_ {ee} $, is smaller than the electron-phonon energy relaxation time $ \tau_e $. In pure
metals for $ T_e \ll \Theta_{D} $, this inequality holds if $ T_e\lesssim k_B \Theta_D^2 /
\varepsilon_F \sim 1 \, $ K. In dirty films, the condition $ \tau_{ee} <\tau_{e} $ is satisfied in
a wider region $ T_e \lesssim 10 $ \, K. \cite{Gershenzon 1982} At high temperatures, the
electrons will be thermalized if their temperature $ T_e> \Theta_{D} (\varepsilon_F / k_B
\Theta_{D})^{1/3} $, that is, $ T_e \gtrsim 10^3 \, $~K.

Unlike $ T_e $, the conditions necessary for introducing the
magnon temperature are much less stringent.
 The condition necessary for introducing the magnon temperature is
that the magnon-magnon collision frequency is much larger than the collision frequency of magnons
with phonons. For four-magnon processes, the average frequency of magnon-magnon collisions
\cite{AkhiezerSpinWaves}

\begin{equation}\label{390}
\bar\nu_{mm}\sim\frac{\Theta_C}{\hbar}\Bigl (\frac{T}{\Theta_C}\Bigr )^4 .
\end{equation}

The average frequency of magnon-phonon
collisions

\begin{equation}\label{360}
\overline{\nu}_{mp}(T)=18.2\,\nu_0  \Bigl(\frac{\Theta_D}{\Theta_C}\Bigr)^{1/2}\,\Bigl
(\frac{T}{\Theta_D}\Bigr )^{5/2} \exp\Bigl(-\frac{\Theta_{D}^{2}}{4\Theta_C T}\Bigr).
\end{equation}
The value $ (\overline{\nu}_{mp})^{-1} $ is equal to the characteristic time of the energy
relaxation of magnons on phonons $ \tau_{mp} $. \cite{Bezuglyj2018} A comparison  of Eq.
(\ref{390}) with Eq.  (\ref{360}) shows that for $\Theta_D \sim \Theta_C$ (as in YIG) the
magnon-magnon collision frequency is significantly higher than the collision frequency of magnons
with phonons for any $T$ and the magnon temperature can be introduced at all thermostat
temperatures $ T_B $.

 \section*{Acknowledgements}

The research leading to these results has received funding from the European  Union's Horizon 2020
research and innovation program under Marie Sklodowska-Curie Grant Agreement No.~644348 (MagIC).

\end{document}